\pdfoutput=1
\documentclass[10pt,twocolumn]{article}

\usepackage[margin=0.75in]{geometry}
\usepackage{times}

\usepackage{graphicx}
\usepackage{amsmath, amssymb}
\usepackage{url}
\usepackage{hyperref}
\usepackage{listings}
\usepackage[dvipsnames]{xcolor}
\usepackage{tabularx}
\usepackage{multirow}
\usepackage{subcaption}  
\usepackage{enumitem}
\usepackage[most]{tcolorbox}
\usepackage{mdframed}
\usepackage{cleveref}
\usepackage{booktabs}

\graphicspath{{Figures/}}

\newcommand{\custombox}[1]{%
  \vspace{0.25cm}%
  \noindent\fcolorbox{black}{gray!10}{%
    \parbox{\dimexpr\linewidth-2\fboxsep-2\fboxrule}{#1}%
  }%
}

\lstdefinelanguage{LLVMList}{
    morekeywords=[1]{default},
    morekeywords=[2]{allow},
    morekeywords=[3]{function},
    morekeywords=[4]{forbid},
    sensitive=true,
    morecomment=[l]{\#},
}

\title{\textbf{Assessing the Feasibility of Selective Instrumentation for Runtime Code Coverage in Large C++ Game Engines}}

\author{
Ian Gauk\\
ASGAARD Lab, University of Alberta\\
\texttt{igauk@ualberta.ca}
\and
Doriane Olewicki\\
Ubisoft La Forge\\
\texttt{doriane.olewicki@ubisoft.com}
\and
Joshua Romoff\\
Ubisoft La Forge\\
\texttt{joshua.romoff@ubisoft.com}
\and
Cor-Paul Bezemer\\
ASGAARD Lab, University of Alberta\\
\texttt{bezemer@ualberta.ca}
}

\date{} 

\begin{document}
\maketitle

\begin{abstract}
Code coverage is a valuable guide for testing, but in AAA games the overhead of instrumentation conflicts with strict performance requirements and can destabilize automated tests.
We propose and assess a selective instrumentation approach tailored to large game engines written in \texttt{C++}, which reduces the scope of instrumentation while preserving relevant coverage data to developer commits.
Our framework integrates into an industrial game testing pipeline, enabling developers to receive immediate coverage feedback on tests run against their changes.
The compilation overhead of our approach is minimal, allowing instrumentation of over 2,000 commits before doubling build time.
In performance evaluations, even the worst-case scenario maintains frame rates above 50\% of the non-instrumented baseline.  
Across two production test suites maintained by our industry partner, our framework caused no automated test failures, avoiding the instability observed under full instrumentation.
Our work shows that commit-level or build-level coverage of large \texttt{C++} game engines can be achieved with minimal overhead and without compromising test stability.
\end{abstract}

\section{Introduction}
Code coverage measures which parts of a program’s source code are executed during testing, and is widely used to identify untested and potentially error-prone code~\cite{piwowarski1993coverage, ivankovic2019code}.
In video game quality assurance, automated in-engine tests and manual playtesting validate the emergent behaviors of tightly coupled and real-time systems~\cite{coppola2024measure, ullmann2023visualising, vsmid2017comparison, lewis2011whats, politowski2021survey}.
For example, testers may evaluate how a destruction system responds to weapon damage either manually or through automated tests.
Without visibility into coverage, identifying untested areas~\cite{piwowarski1993coverage, ivankovic2019code}, and prioritizing testing efforts~\cite{Pan2021MLTSP, jabbar2022test2vec, huang2020regression} becomes nearly impossible.
However, collecting coverage data in AAA games is difficult due to the overhead from collecting code coverage.

Collecting code coverage typically relies on code instrumentation, which introduces significant runtime overhead~\cite{unity2025coverage, ivankovic2019code}.
Although the cost of instrumenting an individual function call is small, the cumulative overhead on frequently executed low-level functions becomes substantial.
Further, as we demonstrate in this paper,  instrumentation overhead can also trigger failures in automated testing pipelines, where tests are subject to timeouts, timing-sensitive assertions, and physics simulations that are sensitive to timing delays.

Coverage support in commercial game engines is inconsistent and often insufficient for large-scale projects.
Unity provides a coverage package integrated with its test framework, but it incurs significant runtime overhead and requires manual annotations to exclude functions~\cite{unity2025coverage}.
Unreal Engine, another widely used \texttt{C++} engine, offers no built-in coverage functionality~\cite{galeone2022unreal}.
These limitations highlight the need for a instrumentation strategy that scales to large game codebases and integrates seamlessly into existing development pipelines.

To address this limitation, we propose and assess a selective instrumentation framework tailored to large-scale games written in \texttt{C++}.
Our framework allows us to limit instrumentation to recently added or modified functions, reducing overhead while preserving coverage information attributable to a commit.
This paper makes the following contributions:  
\begin{enumerate}
    \item A selective instrumentation framework for lightweight runtime coverage collection in large game engines written in \texttt{C++}.  
    \item An empirical evaluation of the impact of different instrumentation levels on build time and runtime performance in a AAA game.  
    \item A qualitative analysis of instrumentation-induced failures in production test suites, describing common failure modes such as test timeouts, timing-sensitive assertions, and physics simulations. 
\end{enumerate}

The remainder of this paper is organized as follows:
\Cref{sec:background} reviews important background concepts and terminology.
\Cref{sec:related_work} reviews related work.
\Cref{sec:framework_overview} describes the architecture and design constraints of our framework.
\Cref{sec:technical_impl} details its integration into a AAA game testing pipeline.
\Cref{sec:experimental_setup} describes the experiments we run to assess our framework and the impacts of instrumentation.
\Cref{sec:threats} discusses threats to validity for our framework and evaluations.
Finally, \Cref{sec:conclusion} summarizes contributions and outlines future applications.

\section{Background}
\label{sec:background}
Understanding and testing software systems, in particular video games, requires specialized tools and techniques.  
In this section, we provide background information on code coverage, dynamic analysis, instrumentation challenges specific to game development, and LLVM instrumentation.  

\subsection{Code Coverage}
Code coverage metrics quantify how much of a program’s source code is executed at runtime.
Common metrics include statement, branch, and path coverage~\cite{zhu1997software}.
Coverage data is typically collected through dynamic instrumentation~\cite{ball1999concept, larus1994instrumentation}.

Coverage metrics are widely used in software engineering to assess test suite completeness, support test prioritization, and guide automated test generation~\cite{Pan2021MLTSP, jabbar2022test2vec, huang2020regression, tang2024chatgpt}.
In practice, coverage also assists with debugging and fault localization~\cite{wong2010family}, may be enforced in CI quality gates~\cite{sonarqube_quality}, and is required in safety-critical domains such as automotive and aerospace software~\cite{nasaSWE219}.
These established uses make coverage an appealing tool for improving testing in games.

\subsection{Dynamic Analysis and Instrumentation}
Dynamic analysis refers to techniques for observing program behavior during execution, such as performance profiling or coverage analysis~\cite{ball1999concept}.  
Code instrumentation is a common method to enable dynamic analysis, involving the insertion of additional instructions into executables~\cite{larus1994instrumentation}.  
Instrumentation can occur at compile time (compiler-based) or at runtime (binary instrumentation).
Compiler-based tools, such as GCC’s \texttt{gcov}~\cite{gccManual} or LLVM Clang~\cite{clangUsersManual}, provide accurate and portable coverage with relatively low overhead.  
Binary instrumentation tools, such as Intel PIN~\cite{intelPIN}, allow coverage to be added post-build but often incur higher runtime costs.  

\subsection{Instrumentation Challenges for Game Engines}
Traditional coverage workflows assume that instrumenting an entire project or module is feasible.
In game engines, however, in-engine tests often activate systems far beyond the intended test scope, producing large amounts of uninformative coverage data and incurring significant runtime overhead.
When the goal is to determine whether specific code changes are exercised by tests, a narrower coverage scope is more effective.
We therefore focus on commit-level coverage, which reduces instrumentation costs while providing actionable feedback to developers.
This scoped approach is particularly well suited to large \texttt{C++} game projects, where full instrumentation is often impractical.

Modern game engines consist of tens of thousands of files and numerous subsystems that execute under strict real-time constraints~\cite{vsmid2017comparison}.
Some low-level functions invoked during rendering or physics loops may be called thousands of times per second.
Instrumenting these hottest functions introduces significant overhead, which can drastically reduce frame rates and disrupt the timing-sensitive nature of the engine.

Typically, game performance is measured using \textit{frame rates}, which measure how quickly the engine can render consecutive images (frames) to the player.
Maintaining a high frame rate is critical because even small drops negatively impact responsiveness and player experience~\cite{madhusudana2021subjective, claypool2006effects}.
In practice, full-scale instrumentation is infeasible for AAA games because:
\begin{itemize}
    \item Instrumenting frequently called functions increases overhead, reducing frame rates, making manual tests cumbersome or impossible.
    \item Overhead impacts all systems across the engine, from system initialization, and physics simulations, which can cause automated tests to fail.
\end{itemize}
These challenges make selective instrumentation necessary to enable code coverage collection in large-scale games.

\subsection{LLVM Instrumentation}
\begin{figure}[tb]
    \centering
    \includegraphics[width=0.95\linewidth]{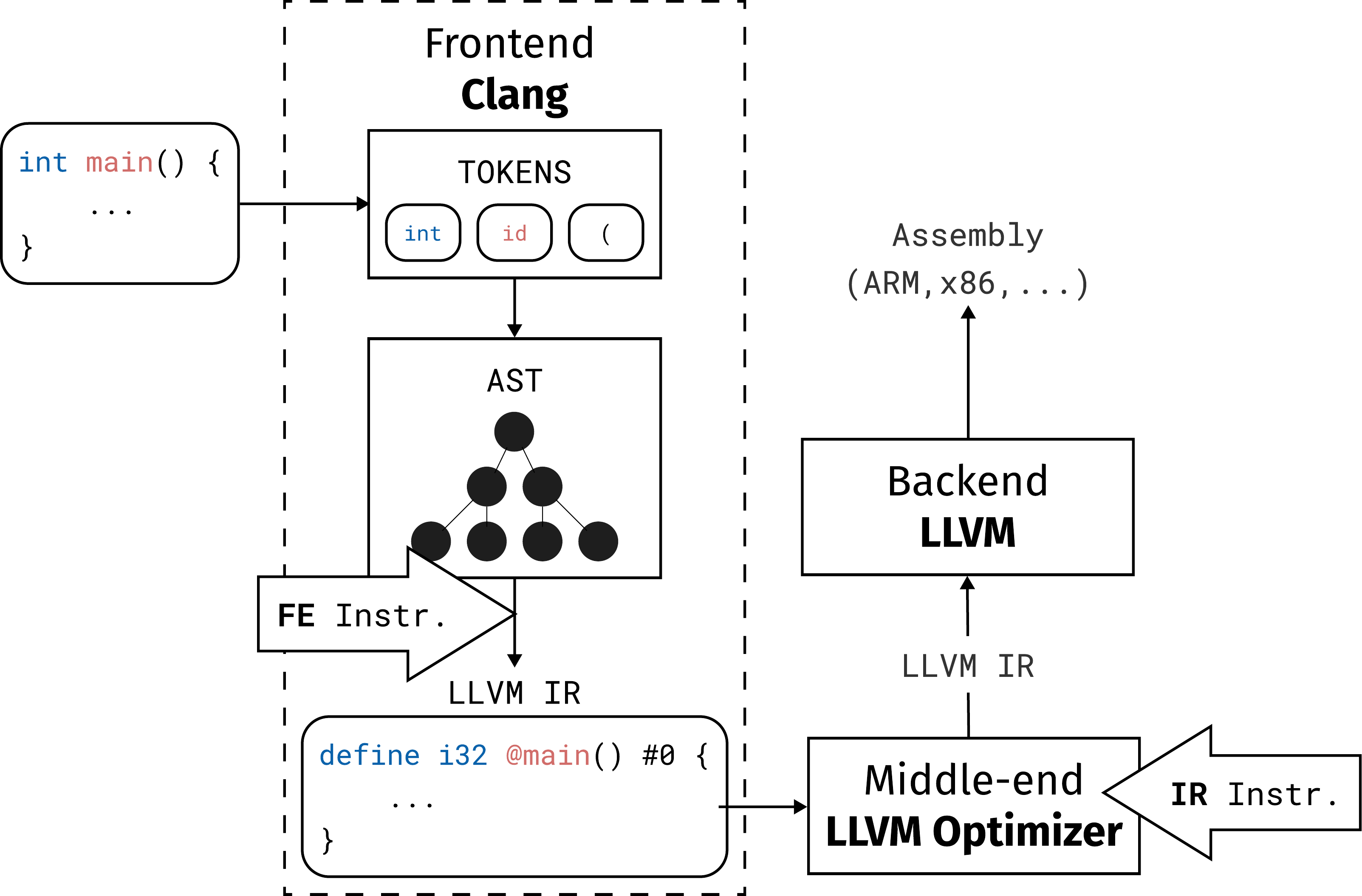}
    \caption{Different types of instrumentation are added at different points in the compilation process, either before or during optimization.}
    \label{fig:instrumentation_types}
\end{figure}

The game engines used at our industry partner are compiled using LLVM Clang,\footnote{A custom fork of LLVM~18 was used.} which provides built-in support for compiler instrumentation passes.  

Clang supports two main types of instrumentation~\cite{clangUsersManual}, each operating at a different stage in the compilation pipeline.  
Clang first parses \verb|C| or \verb|C++| source code into an abstract syntax tree (AST).  
The Clang frontend translates the AST into LLVM’s intermediate representation (IR), a low-level, language-agnostic representation which the LLVM backend can analyze, optimize, and convert into assembly code.  
Instrumentation can be inserted either during or after this AST-to-IR lowering step (see \Cref{fig:instrumentation_types}):  

\begin{enumerate}
    \item \textbf{Frontend-based (FE) instrumentation} inserts counters into the IR during the Clang frontend stage, before any major IR-level optimizations (such as inlining or loop unrolling) are performed.  
    This approach preserves source-level context, enabling precise mapping of counters back to source code.  
    However, because the instrumentation is inserted before optimization, it often results in a higher number of counters and restrict the set of optimizations that the LLVM backend can safely perform, both of which contribute to a higher runtime overhead.  
    Line-coverage using FE instrumentation can be activated using both of the following compiler flags:  
    \begin{itemize}
        \item \verb|-fprofile-instr-generate|: Activates FE instrumentation.  
        \item \verb|-fcoverage-mapping|: Enables line-based coverage through additional mapping information that links source ranges and counters, embedded directly into the IR.  
    \end{itemize}
    
    \item \textbf{LLVM Intermediate Representation (IR) instrumentation} adds counters later in the pipeline, during LLVM’s optimization passes.  
    By this point, many source constructs may have been transformed or removed (e.g., due to inlining), making precise source mapping impractical.  
    As a result, IR instrumentation does not support line-level coverage, but it enables efficient function-level coverage and frequency profiling with significantly lower runtime overhead.  
    IR instrumentation can be activated using \verb|-fprofile-generate|. 
\end{enumerate}

LLVM supports selective instrumentation in both contexts, through the \verb|-fprofile-list=<path>| flag, which takes a \textit{profile list}.
A profile list is a text file that specifies which functions or files should be included or excluded from instrumentation (see~\Cref{lst:instrumentation_list}).
This mechanism allows fine-grained control over instrumentation without modifying the source code.

\section{Related Work}
\label{sec:related_work}
\subsection{Selective Instrumentation}
Selective instrumentation is more commonly used in the high-performance computing (HPC) community to reduce instrumentation overhead while preserving useful information~\cite{kreutzer2022compiler}.
Two main ideas from this line of work are particularly relevant to our approach:

\begin{enumerate}
    \item \textbf{Manual filters in compiler-based instrumentation.}\\
    Performance tools such as \emph{Score-P}~\cite{mey2011score} and \emph{TAU}~\cite{tau-selective} allow users to define include/exclude rules for files, functions, or regions.
    Similarly, LLVM-based coverage instrumentation supports user-defined filters~\cite{clangUsersManual}.
    These filters are commonly used to reduce overhead by skipping hot functions, but require ongoing manual maintenance to remain effective~\cite{kreutzer2022compiler}.
    In contrast, we propose to use them to target functions of interest (e.g., those changed during a commit) for code coverage.
    
    \item \textbf{Automatic filter generation.}\\
    Frameworks like \emph{InstRO}~\cite{instro-site} and \emph{CaPI}~\cite{kreutzer2022compiler} automate filter creation using static analysis (e.g., call-graph traversal).
    Our approach follows this principle but targets code coverage. We automatically build \textbf{selective instrumentation contexts} (SIC)s from commits by analyzing the AST, focusing only on code affected by a commit.
    This minimizes developer effort while controlling overhead and preserving relevant coverage information.
\end{enumerate}

\subsection{Automated QA and Coverage in Video Games}
Recent work on automated testing for video games has leveraged machine learning to explore game states~\cite{albaghajati2020video, gordillo2021improving} and automatically detect bugs~\cite{reza2020video, taesiri2024searching, wilkins2022world}.
These approaches track high-level metrics such as the number of discovered bugs, gameplay coverage, or performance characteristics~\cite{coppola2024measure}.
While effective for exploration and fault detection, they do not measure which parts of the source code are executed.
As a result, prior work cannot identify untested or unreachable code regions, which is essential for assessing test completeness.
Existing studies on game test coverage propose alternative metrics~\cite{coppola2024measure}, but do not examine the feasibility of collecting code coverage in real-time engines.
Our selective instrumentation framework addresses this gap by enabling efficient code coverage collection in AAA games.

\section{Our Framework for Selective Instrumentation}
\label{sec:framework_overview}
The primary goal of our framework is to enable runtime coverage for AAA games by reducing performance overhead using selective instrumentation.
In this section, we discuss our definition of a selective instrumentation context and describe the components that implement selective coverage at the commit level.

\subsection{Selective Instrumentation Contexts}
We adopt the concept of SICs from the high-performance computing community~\cite{instro-site, kreutzer2022compiler}, which limit instrumentation to subsets of functions.
This approach reduces performance overhead while preserving meaningful coverage information.

A SIC represents a subset of functions considered for instrumentation.
The scope of a SIC can vary: it may correspond to a single commit, a batch of commits, a well-defined feature, or even the entire project.
Formally, given the set of all functions in the compilation path $\mathcal{F}$, a SIC is defined as a subset:
\[
\text{SIC} \subseteq \mathcal{F}.
\]

In our industrial setting, we focus on a commit as our SIC.
A commit represents a concrete change in the version control system (VCS), such as adding a feature, fixing a bug, or reverting previous work.
By structuring instrumentation around commits, we align coverage collection with the developer’s immediate context, instrumenting only the functions modified in that commit and avoiding unnecessary overhead.
These commit-scoped SICs also align well with manual tests written by developers, which target recent changes.

\subsection{Framework Components}
To integrate selective instrumentation seamlessly into large-scale \texttt{C++} projects, our framework builds on existing infrastructure and introduces one new component.

\textbf{Instrumentation API.} We introduce an API that exposes methods for generating selective instrumentation files, parsing raw instrumentation data, and storing results in the cloud.
This API integrates with an existing \verb|.NET| application responsible for build orchestration.

To support selective instrumentation end-to-end, we extended the following components:
\begin{enumerate}
  \item \textbf{CI/CD Build Pipeline:} Orchestrates multi-stage builds, extended to trigger selective instrumentation steps as part of the compilation process.
  \item \textbf{Game Wrapper:} A lightweight launcher that manages runtime arguments and environment setup, extended to send profiling information to the instrumentation API on program exit.
  \item \textbf{Database:} Updated to store instrumentation results, with a pipeline to add coverage data to test reports.
\end{enumerate}

\begin{figure*}[tb]
    \centering
    \includegraphics[width=0.95\linewidth]{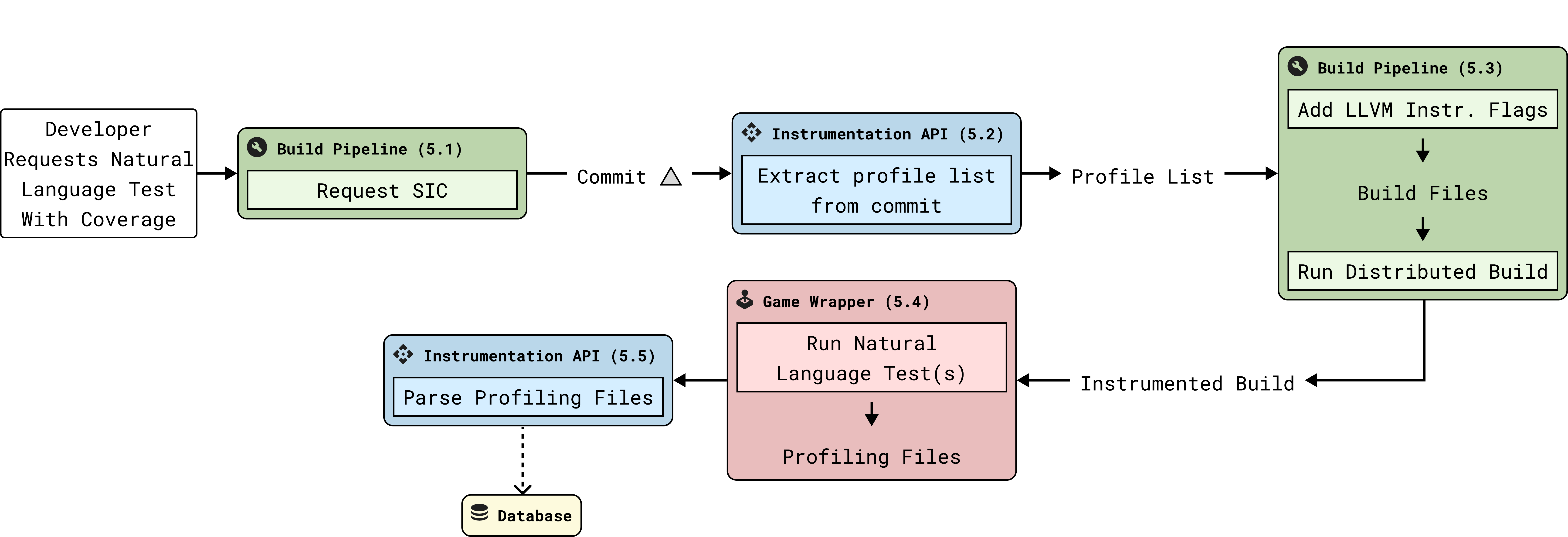}
    \caption{Framework overview}
    \label{fig:framework_overview}
\end{figure*}

\section{Technical Implementation}
\label{sec:technical_impl}
In this section we discuss the specifics of the technical implementation of our framework, and its integration into a custom build and test flow at our industry partner.  
The high-level flow diagram is shown in \Cref{fig:framework_overview}. 

\subsection{Requesting a Build with Coverage}
At our industrial partner, isolated builds can be triggered with specific commits applied on top of an approved baseline build.
These builds are validated by automated tests, and often specify manual test cases written in natural language to test changes, as is standard practice in industry~\cite{viggiato2022similar, viggiato2022nlp}.
This type of build aims to confirm correctness and detect regressions in isolation.

With our framework, when triggering a build for their commit, developers may choose to enable code coverage.
When selected, our custom build pipeline is triggered, which passes the commit to the instrumentation API to extract the SIC.

\subsection{Extracting a Profile List from a Commit}
\begin{figure}[tb]
    \centering
    \includegraphics[width=0.98\linewidth]{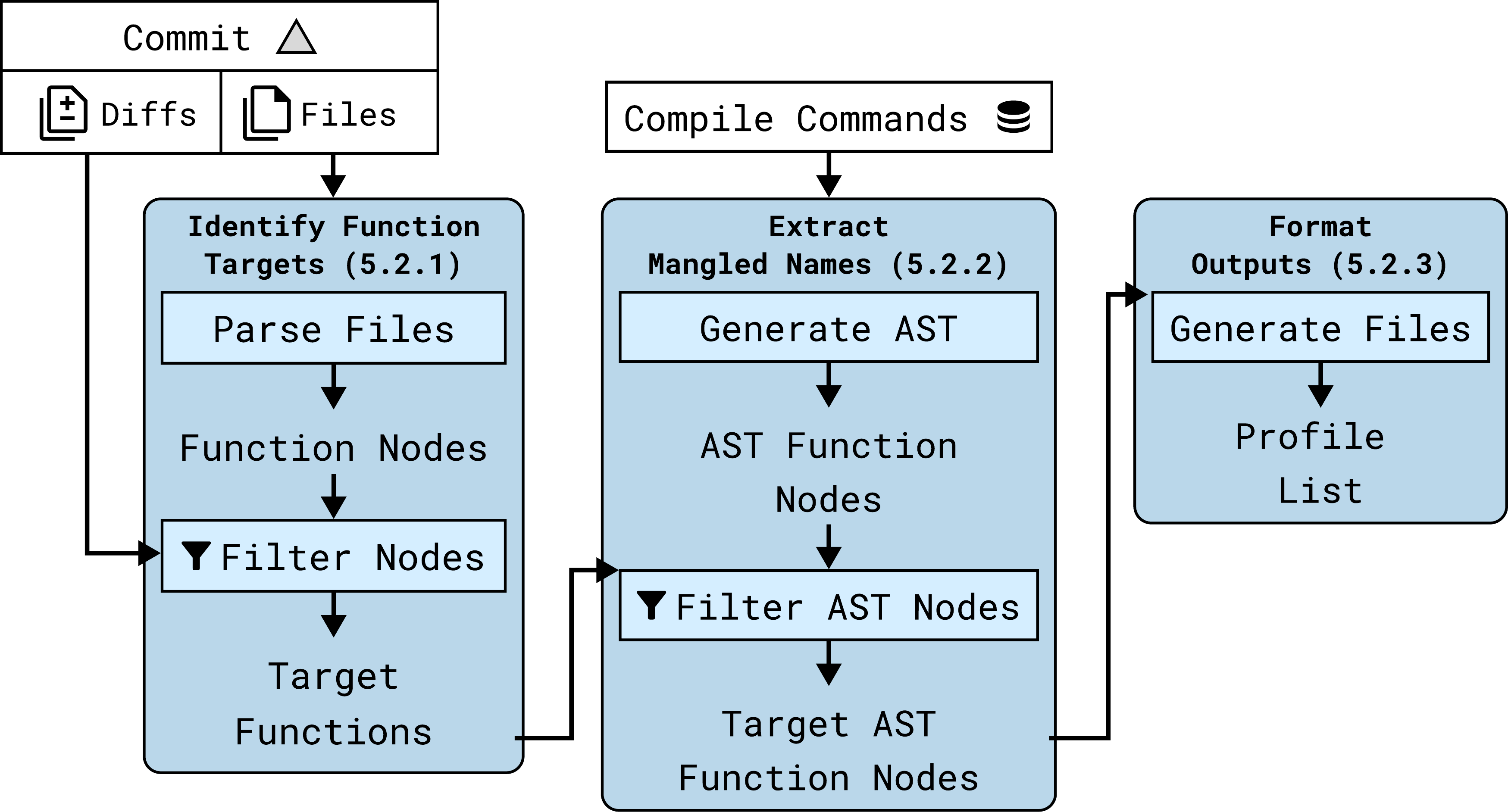}
    \caption{Extracting a profile list from a commit}
    \label{fig:sic_extraction_overview}
\end{figure}

From the commit $\mathcal{C}$ that triggered the build, we generate a profile list to instrument only the functions modified in the commit.
This process unfolds in several steps (see~\Cref{fig:sic_extraction_overview}).

\subsubsection{Identifying Function Targets}
To instrument functions that were modified, first we retrieve the files that were affected by the commit.
Given the unique identifier for $\mathcal{C}$, we retrieve all modified, and added \texttt{.cpp} files for that commit from our VCS.

We identify instrumentation targets by analyzing diffs from the version control system to identify which functions have changed in $\mathcal{C}$.
Using a lightweight parser, we extract the source span of each function declared in the added or modified files.
A function is marked as a target if its source span overlaps with any region reported in the diff.
This rule covers edits occurring inside a function as well as edits that encompass one or more entire function bodies.
All functions declared in newly added files are automatically marked as targets.

At this stage, however, we cannot directly use source-level function signatures to construct the profile list.  
The compiler cannot reliably locate functions from source code signatures, since types may not be fully resolved (e.g., a \texttt{Health} alias for \texttt{float}).  
To ensure unambiguous identification, we must instead use the identifiers that the compiler itself relies on to retrieve function declarations.  
These identifiers are \textit{mangled names}~\cite{msvcDecoratedNames}, which encode complete function signatures after the AST is constructed. 
For each identified target function, we therefore extract the mangled name.

\subsubsection{Extracting Mangled Names}
Mangled names can be generated from the reconstructed AST of the modified source files, which we obtain using the Clang frontend without triggering a full compilation.  
To ensure the AST matches with the actual build context, we rely on a compilation database~\cite{clangJsonCompilationDatabase} which records the exact command used to produce each translation unit.
From this database, we extract the compilation commands associated with the modified file paths and translate their include paths, compiler flags, and macro definitions into a format compatible with our \verb|C#| Clang-based AST generator, \verb|CppAST.NET|~\cite{cppastdotnet}. 
This allows us to reproduce the same compilation environment and reconstruct the AST accurately for mangled name extraction.

From the reconstructed AST, we recursively extract all function nodes originating in the relevant source files.  
To obtain mangled names, we extend the \texttt{CppFunction} node class with a property populated through a P/Invoke call~\cite{dotnet_pinvoke} to Clang's \texttt{ASTNameGenerator::\allowbreak getAllManglings}.  
We then retain only the mangled names for function nodes whose name matches one of the previously identified targets.  
Because the AST fully resolves parameter and return types, we match on the function name alone.  
As a result, overloaded functions may be included even if only one variant was modified, slightly reducing the precision of selective instrumentation.  

\subsubsection{Formatting Instrumentation Outputs}
Mangled names for the target functions modified in $\mathcal{C}$ are formatted into the profile list (e.g., \Cref{lst:instrumentation_list}), where instrumentation is disabled by default and explicitly enabled only for the selected functions.

\begin{lstlisting}[
caption={
Example of a selective instrumentation \texttt{.list} file.
Function level specifications are specified using mangled names
},
label={lst:instrumentation_list},
float,
breaklines=true
]
# Selective Instrumentation List
function:?ApplyForces@Physics@AEXV3@Z=allow
function:?ComputePath@AI@AEXGraph@Z=allow
\end{lstlisting}

\subsection{Creating the Game Package}
Once the instrumentation API returns the profile list, instrumentation flags can be added (see~\Cref{sec:background}).
The profile list is referenced in all relevant build files generated by the build system.
A distributed build is then triggered across multiple workers, and the instrumented game package is ran through automated validation tests and published to testers.
The build system generates a unique build identifier for each build, which can be used to link coverage data with its corresponding build.

\subsection{Running Tests on an Instrumented Game Package}
Once an instrumented build is produced it can be handed off to the QA team and executed like any other test build.  
Instrumented builds automatically produce raw coverage files, generated upon exiting the game.  
We added an exit hook to the game wrapper which forwards these files to the instrumentation API for parsing and generating reports.

\subsection{Creating Coverage Reports}
The instrumentation API uses \verb|llvm-profdata|\footnote{\scriptsize\url{https://llvm.org/docs/CommandGuide/llvm-profdata.html}} and \verb|llvm-cov|\footnote{\scriptsize\url{https://llvm.org/docs/CommandGuide/llvm-cov.html}} to process raw coverage data.  
Coverage information is uploaded to a database associated with the corresponding build identifier.  
Coverage reports are then linked directly in the test report, providing developers with feedback on which functions were covered during testing.

\section{Experiments}
\label{sec:experimental_setup}
We measured the performance impacts of instrumentation along multiple axes.
In particular, we assessed how instrumentation impacts compilation performance, runtime performance and test performance. 
The following sections describe the experiments we ran to understand the trade-offs between the number of functions instrumented in terms of compilation time (\Cref{sec:compile_time}), runtime performance (\Cref{sec:performance}), and automated test failures (\Cref{sec:automated_tests}).

\newcommand{\IFR}{$\mathrm{IFR}$}
Throughout our experiments, to quantify the extent of selective instrumentation in a build, we use the \textit{Instrumented Function Ratio (\IFR)}.
\IFR{} measures the proportion of the codebase that is instrumented:
$$
    \mathrm{IFR} = \frac{|\mathrm{SIC}|}{|\mathcal{F}|}
$$
For instance, if \IFR{}$=0.5$, half of the codebase is instrumented.
Note that the games we consider at our industrial partner each have more than 1 million functions.

\subsection{Experiment 1: Compilation Time}
\label{sec:compile_time}
Selectively instrumenting code can affect the time taken to compile the game.
In this first experiment, we investigated to what extent adding selective instrumentation increases compilation time.

We considered compilation time in particular, which refers to the duration required to convert the project code into a runnable form.
In contrast, a complete build includes additional steps such as downloading tool dependencies and processing data assets, which are unrelated to code instrumentation.
Keeping compilation time low is important in practice, as it directly influences the overall throughput of pre-merge testing, and the developer iteration speed.
In our framework, compilation time overhead arises from two main sources:
(1)~the prerequisite cost of extracting the profile list from a commit, and
(2)~the additional cost of selectively adding instrumentation during compilation.

\subsubsection{Profile List Extraction Overhead}
\newcommand{\extractionTime}{$\mathrm{PER}$}
\newcommand{\extractionTimePerFile}{\extractionTime$/\mathrm{file}$}
To evaluate the cost of extracting profile lists from commits, we measured the additional time required to compile builds that include profile list extraction.

\textit{Experimental Setup.}
We extracted profile lists for 100 recent commits and we report the additional wall-clock time relative to a baseline.
As the latency of profile list extraction directly contributes to developer perceived build time, wall-clock time is the most informative metric.
The baseline corresponds to the average build time over five compilations without profile extraction.

To report this cost, we define the \textit{Profile Extraction Ratio} (\extractionTime), which quantifies the proportional increase in wall-clock time relative to the baseline:  
$$
{\mathrm{PER}} = \frac{\text{Profile List Extraction Time}}{\text{Baseline Wall-Clock Time}}
$$
For example, \extractionTime{} $= 0.5$ indicates that profile extraction added 50\% of the baseline compilation time to the overall build time.  
Because commits can vary significantly in scope, we also report the extraction time per file, \extractionTimePerFile{}, to normalize across commits.
We normalize by files and not functions because the dominant cost comes from AST generation, which operates at a file level.

\begin{figure}[tb]
    \centering
    \subfloat[
    Profile list extraction time for 100 commits (Min: 11.31\%, Median: 17.56\%, Max: 183.61\%).]
    {
        \includegraphics[width=0.80\linewidth]{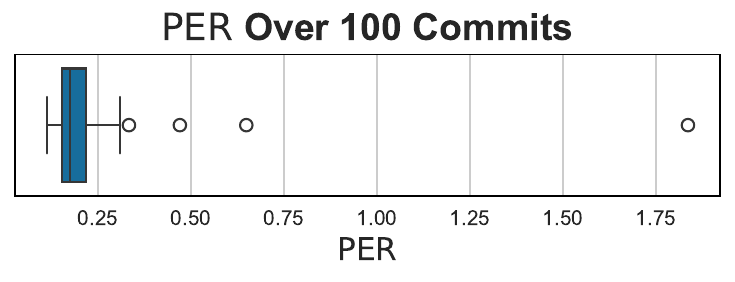}
        \label{fig:wall_time}
    }

    \subfloat[Profile list extraction time for 100 commits, normalized by the number of files changed (Min: 0.75\%, Median: 8.33\%, Max: 25.59\%).]
    {
        \includegraphics[width=0.80\linewidth]{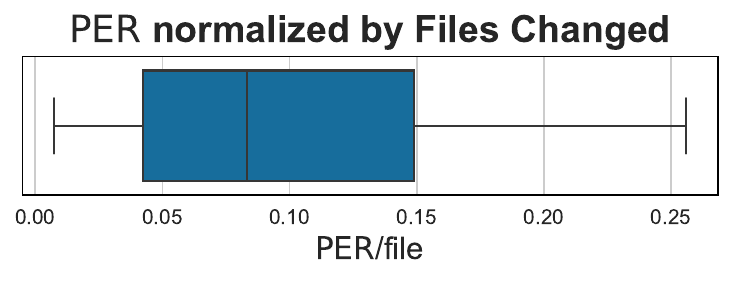}
        \label{fig:compilation_time}
    }

    \caption{Overhead introduced by profile list extraction}
    \label{fig:extraction_overhead}
\end{figure}

\textit{Results.}
\textbf{Profile list extraction is fast relative to compilation, with only rare large commits causing significant slowdowns.}
As shown in \Cref{fig:wall_time}, a median commit increased the compilation time by 17.56\%, while the largest commit resulted in a compilation 183.32\% longer than the baseline.
The outlier was a refactoring change involving 98 files and 799 functions.
Such commits are uncommon, but they illustrate how extraction cost scales with the size of a commit.

\textbf{Profile list extraction overhead is proportionate with the number of files modified in a commit.}
AST generation dominates the extraction cost, so commits that modify many files, or files with many functions, classes, or with large header hierarchies incur higher overhead due to the resulting larger ASTs.
As shown in \Cref{fig:compilation_time}, the number of files changed in a commit is a reliable predictor of extraction time, and our results include no extreme outliers in this distribution.

\custombox{
    The median overhead of extracting profile lists from the last 100 commits submitted to a AAA game added 17.56\% to the baseline compilation time.
    We found that the number of files changed in a commit is a reasonable heuristic to estimate the profile extraction overhead.
}

\subsubsection{Compilation Overhead}
\newcommand{\cpuTime}{$t_{\mathrm{CPU}}$}
To understand how selective instrumentation affects compilation cost, we measured the total CPU time used across compilations with varying \IFR. 

\textit{Experimental Setup.}
To explore the scalability limits of our approach, we compiled builds that instrumented increasingly large fractions of the codebase, up to 50\%, in addition to the commit-based SICs (see \Cref{tab:perf_builds}).
These large \IFR{} configurations do not reflect realistic instrumentation scenarios, as even 1\% of a million-function codebase already targets a substantial number of functions, but they allow us to observe trends under stress conditions.

Since compilations are distributed across multiple workers, wall-clock time is sensitive to queueing delays, resource contention, and machine-level variability.
In contrast, total CPU time aggregates compute usage across all workers, providing a stable measure of the computation overhead from instrumentation.
We define the \textit{Total CPU Time Ratio} (\cpuTime) to quantify this overhead:  
$$
t_{\mathrm{CPU}} = \frac{\text{Total CPU Time}}{\text{Baseline CPU Time}}
$$ 
For example, \cpuTime{} $= 4$ indicates that the instrumented build consumed four times more CPU time than the baseline.  
The baseline for \cpuTime{} is computed in the same way as for wall-clock measurements and corresponds to the average total CPU time across five non-instrumented builds.
This metric allows us to compare the scalability of instrumentation across different build sizes and contexts.

\begin{figure}[tb!]
    \centering
    \subfloat[\cpuTime{} scales linearly with \IFR{} for both FE and IR instrumentation. We calculate the least-squares fit using \texttt{scipy}, with the default 95\% confidence interval.]
    {
        \includegraphics[width=0.90\linewidth]{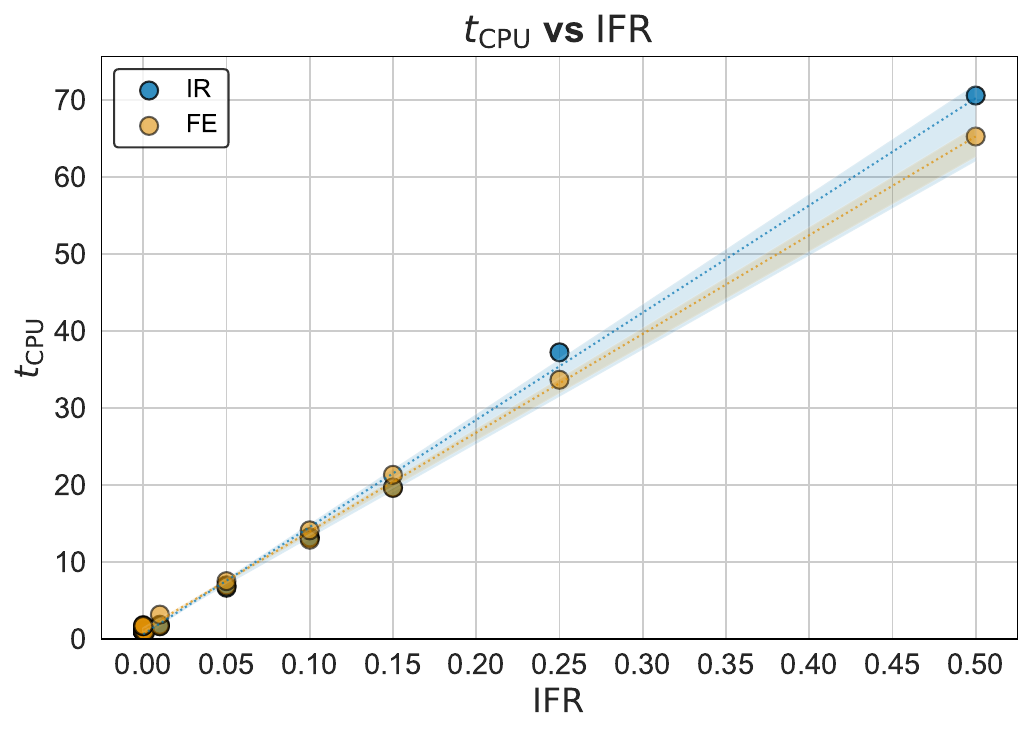}
        \label{fig:cpu_scaling}
    }

   \subfloat[(Crop of \Cref{fig:cpu_scaling}) Compilation time remains less than twice the baseline until a certain \IFR{} threshold at around 1\%.
   Median, largest and batch refer to SICs extracted from the last 100 commits to a AAA game, seen in \Cref{tab:perf_builds}]
   {
        \includegraphics[width=0.95\linewidth]{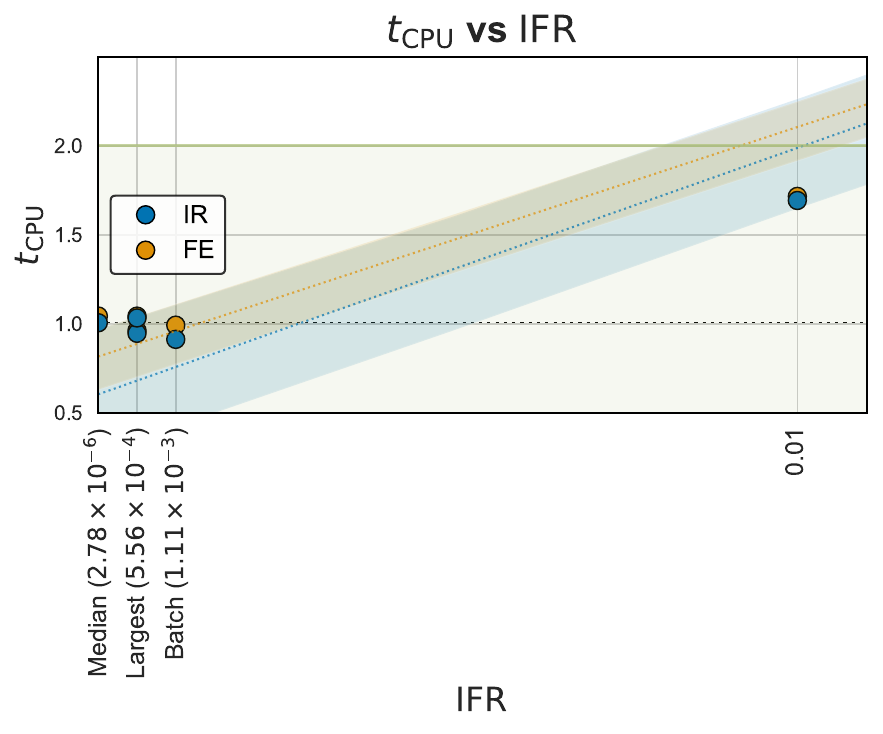}
        \label{fig:maximum_ifr}
    }
    
    \caption{Relationship between \IFR{} and \cpuTime{} at different scales.}
    \label{fig:ibr_effect}
\end{figure}

\textit{Results.}
\textbf{\cpuTime{} scales linearly and quickly with \IFR{} for selective instrumentation.}
As seen in~\Cref{fig:cpu_scaling}, the steep linear regression slope values (128.08 for FE and 139.84 for IR) indicate that compilation time increases significantly with \IFR.
Even modest increases in profile list size can lead to substantial increases in CPU build time.
Since fully instrumented builds (i.e., without selective instrumentation) have low \cpuTime{}~(1.75 and 1.00 for FE and IR instrumentation respectively), we attribute the increased compilation cost to the size of the profile list used in instrumentation.
We hypothesize that the high cost of increasing the profile size stems from the internal design of LLVM selective instrumentation.
Each worker and thread in a parallelized build process must maintain a full copy of the profile list and check this list (stored as a string map) against every function in the translation unit.
This design has compounding memory and CPU overhead, especially as we increase the proportion of functions in our profile.

\textbf{The compilation time increase due to selective instrumentation at low \IFR{} levels is minimal.}
As shown in \Cref{fig:maximum_ifr}, for commit-based SICs (see \Cref{tab:perf_builds}) have compilation times that are effectively indistinguishable from that of the baseline, since only a small fraction of the codebase is instrumented.
In practice, many commits could be instrumented simultaneously before the compilation time doubles.
From the observed trends, we estimate that instrumentation could be budgeted for up to 1\% of the codebase (\~2{,}000 commits) while keeping the build time under twice the baseline.

\custombox{
We found developers can expect compilation to complete in a time similar to a non-instrumented build using our framework for commit-based SIC.
This minimal overhead makes our approach suitable for CI/CD pipelines, where multiple commits are batched into one build.
Extrapolating from the average number of functions modified per commit, we estimate that over 2,000 commits could be instrumented before reaching a $2\times$ \cpuTime{} budget for our application.
}

\subsection{Experiment 2: Runtime Performance Impact}
\label{sec:performance}
\newcommand{\fpsRatio}{$\mathrm{FPS}~\text{Ratio}$}
\newcommand{\FPS}{$\mathrm{FPS}$}
Next, we investigate the runtime performance impact of instrumentation. 

\textit{Experimental Setup.}
Performance tests are often used in industry to detect load time and frame rate regressions due to changes to code and data in game builds.
These tests can quantify changes in performance due to FE and IR instrumentation at different \IFR.

\begin{table}[tb]
\centering
\caption{Selective Instrumentation Contexts}
\begin{tabular}{>{\raggedright\arraybackslash}p{0.65\linewidth} >{\raggedleft\arraybackslash}p{0.25\linewidth}}
\toprule
\textbf{SIC} & \textbf{IFR} \\
\midrule
\textbf{Baseline} No Instrumentation & $0.00$ \\
Fully Instrumented & $1.00$  \\
Median Commit (100 Most Recent) & $2.78 \times 10^{-6}$ \\
Largest Commit (100 Most Recent) & $5.56 \times 10^{-4}$ \\
Batch of Commits (100 Most Recent) & $1.11 \times 10^{-3}$ \\
Worst Case Runtime Performance Commit & $5.56 \times 10^{-4}$ \\
\bottomrule
\end{tabular}
\label{tab:perf_builds}
\end{table}

To assess runtime performance impact, we ran performance tests on realistic instrumentation contexts derived from actual development activity.
Again, we considered the 100 most recent commits to a AAA game at our industrial partner, and from these, we defined four SIC:
\begin{enumerate}
    \item \textit{Median commit:} A SIC from the median commit in terms of number of functions modified.  
    This SIC represents a typical developer change, providing a baseline for the expected overhead in day-to-day development.

    \item \textit{Largest commit:} A SIC from the commit with the highest \IFR{} among the 100 most recent commits.  
    This simulates an unusually large change, allowing us to assess the impact of instrumentation on performance when a commit touches many functions.

    \item \textit{Batch of commits:} A SIC batching all modified functions across the 100 most recent commits.  
    This SIC reflects a scenario where multiple commits are grouped together, introducing changes across many systems.
    It provides insight into how our approach could scale to larger \IFR.

    \item \textit{Worst-case commit:} A SIC containing only the most frequently invoked functions, matching the number of functions in the real largest commit.  
    This scenario mimics a large refactoring that affecting low-level functions most heavily invoked during gameplay.
    We identified these functions by collecting a runtime profile during the performance test, and selecting the top-ranked functions by invocation frequency to form the instrumentation context.
\end{enumerate}

These SIC allow us to evaluate how our approach scales with typical and extreme changes in a production environment.
A summary is shown in \Cref{tab:perf_builds}, with two builds per SIC for each instrumentation type.
We ran performance tests on each build five times on the same hardware to ensure consistency.

The performance test is an automated session which places the player character in a variety of in-game scenarios.
These scenarios are designed to trigger different subsystems of the game engine, such as rendering, physics, asset loading, and AI.
Performance data, such as frame time, is sampled and aggregated across the performance test.

Frame time measures the duration taken to render a single frame.
We convert this to the more intuitive metric, frames per second (\FPS) by taking the inverse of frame time (in seconds).
For example:
$$
\left(\frac{0.016\overline{6}\text{s}}{\text{Frame}}\right)^{-1} = 60\ \mathrm{FPS} 
$$
We retrieve average frame rates from each test run, normalized against the baseline \FPS{}, \textbf{Frame Rate Ratio (\fpsRatio):}
$$
    \mathrm{FPS}~\text{Ratio} = \frac{\text{Average Test \FPS}}{\text{Average Baseline \FPS}}
$$
To contextualize this, an \fpsRatio{} of 0.5 means that the average frame rate is half that of the baseline build, or 30~\FPS{} when the baseline was 60~\FPS.

We report average frame rate values to capture overall performance impact of instrumentation.
Although frame time distributions in games often include outliers from events like asset streaming or garbage collection, our measurement setup samples only instantaneous frame rate values and does not provide direct access to detailed frame-time traces.
As a result, the average frame rate is the most reliable statistic available to summarize overall runtime performance across test runs.

\begin{figure}[tb]
    \centering
    \includegraphics[width=0.95\linewidth]{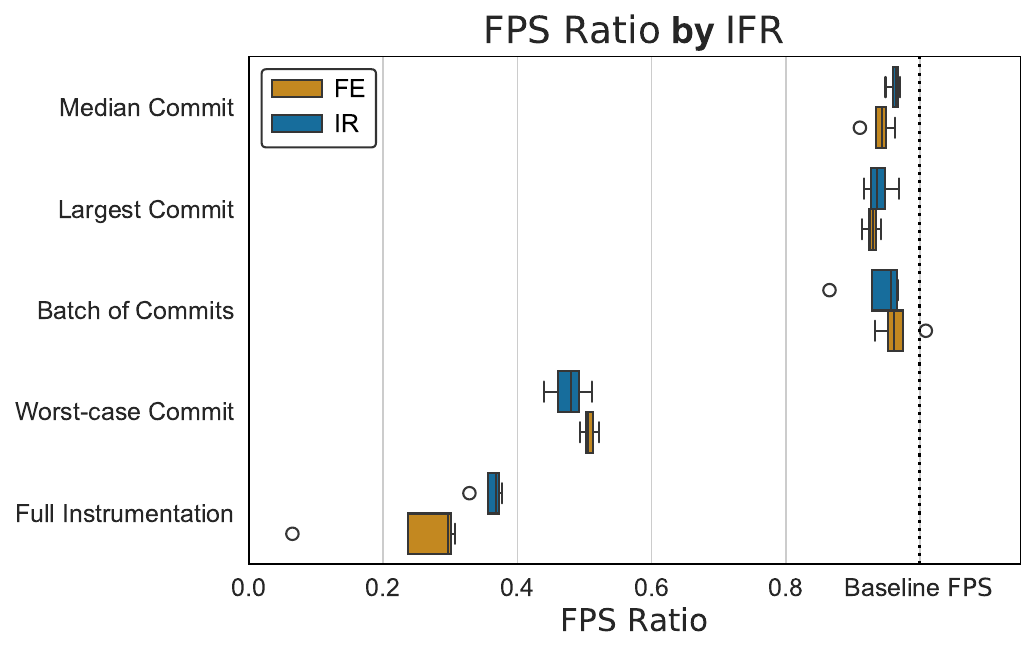}
    \label{fig:fps_avg}
    \caption{\FPS{} ratio after instrumentation}
    \label{fig:performance_overhead}
\end{figure}

\textit{Results.} 
We show the performance impact of different SIC and instrumentation types in \Cref{fig:performance_overhead}.

\textbf{Instrumenting commits causes minimal runtime performance overhead.} 
Builds based on the \textit{median}, \textit{largest}, and \textit{batch of} 100 commits consistently maintained an \fpsRatio{} above 0.9.
This indicates that typical selective instrumentation has a negligible impact on gameplay performance, regardless of whether FE or IR instrumentation is used.

\textbf{High-frequency function instrumentation is costly.}
Targeting only the most frequently called functions for instrumentation in the \textit{worst-case} commit SIC caused \FPS{} to drop to approximately 50\% of the baseline.
This demonstrates that the frequency of calls to instrumented functions has a strong impact on \FPS{} over the number of instrumented functions.
However, we infer that this worst-case scenario is unlikely, considering that the batch of 100 commits had no overlapping functions with the worst-case, and much higher frame rates.

\textbf{When instrumenting the entire codebase, both FE and IR approaches led to substantial performance degradation.}
The median \fpsRatio{} dropped to 0.297 for FE and 0.369 for IR.
In both cases runtime performance was severely affected, while a larger divergence between instrumentation types shows that source mapping costs (FE) become more pronounced when more functions are instrumented.

\custombox{
    The runtime performance impact of instrumentation is highly sensitive to the selection of functions.
    While typical commits show negligible runtime overhead, targeting highest-frequency functions or instrumenting the entire game can drastically reduce \FPS.
}

\subsection{Experiment 3: Automated Test Stability}
\label{sec:automated_tests}
While frame rate drops caused by instrumentation can negatively impact manual testing, it may seem reasonable to fully instrument automated tests, where frame rate does not directly affect the tester experience.
In this experiment, we investigate whether instrumentation affects the stability of automated tests.

\newcommand{\neo}{Game$_A$}
\newcommand{\ovr}{Game$_B$}
\textit{Experimental Setup.}
We assessed test outcomes using approved test suites for two games under production: \neo{} and \ovr{}, to determine the failure patterns induced by instrumentation overhead and evaluate the patterns that emerge across distinct titles and test suites.
These games belong to different genres and have test suites maintained by separate teams, but using the same testing framework.
The \neo{} suite contains 31 tests and the \ovr{} suite includes 44 tests, targeting AI and 3Cs functionality.
All tests pass on a non-instrumented build on the same hardware, so any deviations can be attributed to instrumentation overhead.

To isolate the impact of instrumentation each suite was executed on two fully instrumented builds: one using FE-based instrumentation and one using IR-based instrumentation.
We also ran tests on the commit with the worst-case SIC (see \Cref{tab:perf_builds}) to observe the upper bound of expected selective instrumentation effects.
This design does not cover all possible workflows, but provides a controlled comparison of how different instrumentation modes interact with real automated tests.

\textit{Quantitative  Results.}
\textbf{Full FE instrumentation introduced automated test failures in both games, while even in the worst-case our framework resulted in no test failures.}
Full FE introduced failures in both automated test suites due to its high runtime overhead, causing timeouts or nondeterministic behavior in time-sensitive tests.
Under full FE instrumentation, 23 of 44 of tests failed on \ovr{} and 6 of 31 on \neo.
In contrast, full IR instrumentation produced no test failures across either test suite, indicating that its runtime overhead is low enough to preserve test stability.
IR instrumentation, however, offers less granular coverage information, which limits its suitability for scenarios that require detailed line-level insights.

\textbf{Considering both FE and IR instrumentation for the worst-case SIC resulted in no test failures.}
This demonstrates that selective instrumentation reduces overhead and preserves automated test stability while providing fine-grained coverage.

\textit{Qualitative Results.}
The following failure modes were observed only under full FE instrumentation and illustrate how high runtime overhead can destabilize automated test suites in industrial game engines.

\textbf{Instrumentation caused initialization delays leading to test timeout failures.}
This failure mode is characterized by tests exceeding their allotted time, generally during initialization and before reaching the test-use case.
Timeouts are commonly used in automated tests to detect potential hangs and prevent indefinitely stalled runs in CI/CD pipelines.
All tests across both games were executed under such timeout constrains.
Due to increased initialization times under full FE instrumentation, tests in \ovr{} failed when world loading exceeded default timeouts, while those that passed had longer overridden timeouts.
Although instrumentation increased test duration in both games, we did not observe any timeout-related failures in \neo{}.

To determine whether test timeouts were the only issue, we removed default timeouts in \ovr{} to allow tests to continue past initialization into their test use-case.

\textbf{Instrumentation overhead caused timing-based assertions to fail.}
This failure mode is characterized by timing-sensitive systems running slower than expected, causing statement level test assertion timeouts to fail.
Full FE instrumentation introduced delays in asset loading and in systems such as behavior tree transitions, affecting both games.
These delays caused interactions or scripted events to miss expected time windows, and assertions checking for timely updates to fail.
For example, NPCs expected to deal damage to a player character within a short interaction window loaded too slowly to meet the assertion threshold, and NPCs reacting to a sequence of behaviour transitions responded too late to satisfy test requirements.

\textbf{Instrumentation overhead degraded physics performance, breaking movement-related tests.}
This failure mode is characterized by reduced physics update rates that prevent physics-driven behaviours from reaching expected states within the allotted number of frames.
Failure modes differed across titles: while tests did not fail in \neo{} due to initialization timeouts, several tests failed due to slower-than-normal physics simulation under full FE instrumentation.
Tests that validate movement or physics-driven behaviors in \neo{} failed because the physics engine could not maintain expected update rates.
For example, in a movement test, a character scripted to reach a target velocity of 3.5\,m/s within a fixed number of frames did not accelerate sufficiently, causing the associated assertion to fail.

\custombox{
    Full FE instrumentation provides detailed line-level coverage but introduces significant overhead that destabilizes automated tests, causing failures due to timeouts, time-based assertions, and physics inconsistencies.
    IR instrumentation reduces overhead and avoids failures but offers less granular coverage information.
    Our approach of selective instrumentation preserves test stability while still delivering precise line-level coverage for the parts of the code that matter most.
}

\section{Threats to Validity}
\label{sec:threats}

\textit{Internal validity.}
A threat to internal validity is that selective instrumentation may offer limited benefit when IR-based instrumentation alone is sufficient.
In our experiments, IR instrumentation exhibited consistently low overhead and did not induce test failures, reducing the practical advantage of narrowing the instrumentation scope.
However, IR instrumentation provides less granular coverage information, which limits its usefulness for debugging tasks that require precise line-level insight.
Our selective approach aims to reduce the overhead of FE instrumentation while retaining this fine-grained coverage.

A second threat concerns limitations in how performance data is collected.
Our industry partner’s tooling provides only periodic samples of instantaneous FPS and does not expose per-frame timing information.
As a result, our analysis cannot examine detailed frame-time behavior or quantify worst-case latency, and must rely on average FPS as the most reliable available metric.

A further threat is that restricting instrumentation to functions modified in a commit may overlook the instrumentation of other relevant execution paths.
The implications of this restriction are not fully understood.
Future work will explore broader SIC definitions, such as AST or call-graph based expansion of the instrumentation context.

\textit{External validity.}
An external threat to validity is that Experiments~1 and~2 were conducted on a single AAA game title.
While this provides insight into instrumentation at production scale, it is unclear whether the observed trends in compilation time and runtime performance generalize to other games or codebases.
Differences in engine architecture, build systems, or project scale may lead to different instrumentation overhead characteristics.
Future work should evaluate the proposed framework across multiple titles of varying size and complexity to better assess the generalizability of these results.

Another external threat to validity is that the findings from Experiment~3 are closely tied to the automated test suites and hardware configurations used in our study.
Although these results illustrate how instrumentation can affect test stability, they may not generalize to other projects.
Testing practices vary across teams, including the use of different assertion mechanisms and timeout thresholds, which can lead to distinct failure modes.
In our case, timeouts, commonly used to detect deadlocks or initialization hangs, were the primary cause of failures in FE-instrumented builds.
While higher-end hardware may alleviate some performance-related failures, automated tests are typically expected to remain stable on the lowest supported hardware.
Future studies should therefore examine the effects of selective instrumentation across a broader range of test suites and hardware configurations.

\section{Conclusion}
\label{sec:conclusion}
Collecting code coverage in AAA games is challenging due to the high runtime overhead of full code instrumentation.
We introduced a framework for selective code instrumentation that targets changes at the commit-level, allowing coverage collection with minimal disruption to manual or automated testing.
We further explored the compilation and runtime performance considering different levels and types of instrumentation (Frontend and LLVM Intermediate Representation).
Our evaluation shows that our framework maintains average frame rates above 50\% of the non-instrumented baseline, while avoiding automated test failures.
Further, we show that compilation overhead is minimal in practice, and we could instrument 2,000 commits before doubling compilation time.
These results show that commit-based selective instrumentation integrates seamlessly into a build and testing flow, and is feasible for large \texttt{C++} game engines.

Based on these findings, we recommend selective instrumentation as a practical alternative to full frontend instrumentation in production settings where performance constraints or automated test stability concerns make full instrumentation impractical.
Applying selective instrumentation allows developers to confirm what parts of a commit were invoked during testing, supporting more informed testing decisions by developers and quality assurance teams.
The approach can be deployed in other large game engines written in \texttt{C++} with minimal instrumentation overhead.

\section{Data Availability Statement}
Due to the research being pursued in an industrial closed source environment, information concerning the proprietary code and the actual running time could not be disclosed, leading to the data and source code not being available.

\bibliographystyle{plain}
\bibliography{ref}

\end{document}